\documentstyle[prb,eqsecnum,aps]{revtex}
\begin{document}

\draft
\title{Bohmian Mechanics with Discrete Operators}
\author{R. A. Hyman\cite{aflcio}, Shane A. Caldwell, Edward Dalton}
\address{Department of Physics, DePaul University, Chicago, IL
60614-3504}
\date{\today}
\maketitle

\begin{abstract}
A deterministic Bohmian mechanics for operators with continuous
and discrete spectra is presented. Randomness enters only through
initial conditions.  Operators with discrete spectra are
incorporated into Bohmian mechanics by associating with each
operator a continuous variable in which a finite range of the
continuous variable correspond to the same discrete eigenvalue. In
this way Bohmian mechanics can handle the creation and
annihilation of particles. Examples are given and generalizations
are discussed.
\end{abstract}

\section{introduction}
In 1951 David Bohm introduced an extension of quantum mechanics,
now called Bohmian mechanics, in which particles have definite
positions and velocities at all times, including between
measurements\cite{bohm}.  He also showed how the same idea can be
applied to field configurations.  Bohmian mechanics agrees with
all of the experimental predictions of the Copenhagen
interpretation of quantum mechanics but has the advantages of a
smooth transition to classical mechanics, the absence of
wave-function collapse, and not having to separate the universe
into quantum systems and classical measuring devices. Furthermore,
the dynamics is deterministic and time reversible. Randomness
enters only in the initial conditions of the particle and field
configurations. On the other hand, Bohm constructed his mechanics
in first quantized form which is fine for the nonrelativistic
Schrodinger equation but it can not handle fermion particle
creation and annihilation in the second quantized representation
of quantum field theory.   In this paper we generalize Bohmian
mechanics so that it can handle operators with discrete spectra,
and thereby accommodate second quantized fermionic field theories
while remaining deterministic.

In Bohmian mechanics, the quantum state and Hamiltonian do not
constitute a complete description of physical reality as they do
in the Copenhagen interpretation of quantum mechanics. Bohmian
mechanics singles out a set of dynamical variables associated with
a particular set of commuting operators (for example, the
positions of all of the particles) that are needed in addition to
the quantum state and Hamiltonian to constitute a complete
description of physical reality at all times. J. S. Bell has
coined the term 'beables', short for 'maybeables' for those
operators promoted to reality status\cite{bell}. The hesitancy is
built into the word to indicate, somewhat analogously to gauge
freedom, that the set of commuting operators is not uniquely
determined by experiment.  Bohm's original formulation was
expressed in first quantized notion  His formalism required that
all of the beables have continuous spectra which is fine when
electron position operators are the beables.  However, the Dirac
equation presents a problem in this formalism since it requires an
infinite number of negative energy electrons in the vacuum, so the
most naive extension of Bohmian mechanics to the relativistic
realm requires the calculation of an infinite number of
trajectories even when there is nothing really there. If instead
one thinks of the vacuum as a vacuum, and moves to a second
quantized representation, then one is faced with the possibility
of electron-positron pair creation and annihilation. This can not
be straightforwardly treated with continuous variables alone since
the number of particles is not a continuous variable.

In 1984 J. S. Bell constructed a Bohmian-like mechanics for
beables with discrete spectra, in particular he considered fermion
configurations in relativistic field theories\cite{bell}. He gave
up on particle trajectories altogether and considered instead the
fermion number density operators as the beables. Unlike Bohmian
mechanics, Bell's mechanics is stochastic. Bell was dissatisfied
with this since "...the reversibility of the Schrodinger equation
strongly suggests that quantum mechanics is not fundamentally
stochastic in nature. However I suspect that the stochastic
element introduced here goes away in some sense in the continuum
limit.\cite{bell}"  Since Bell there have been several other
contributions to Bell-like dynamics for relativistic quantum field
theory, some bringing back explicit particle trajectories, but
like Bell regrettably sacrificing determinism\cite{others}. In
this paper we introduce an alternative to Bell's mechanics that is
a deterministic time-reversible Bohmian mechanics for operators
with both discrete and continuous spectra. The determinism and
time reversible invariance is present at the course grained level.
No continuous limit is necessary, and no modification of the
Hamiltonian need be made. Since our extension of Bohmian mechanics
allows for discrete beables it is able to handle particle creation
and annihilation in second quantized field theories, and thereby
dispense with one of the objections to Bohmian mechanics.

In section II of this paper we present a generalization of Bohmian
mechanics for operators with continuous spectra that is also
amenable to operators with discrete spectra.  In section III we
incorporate operators with discrete spectra into the formalism. In
section IV we present the exactly solvable case of Bohmian
mechanics for one beable, and an intriguing visualization of
Bohmian mechanics for any number of beables. In section V contains
a summary of our results.

\section{Bohmian mechanics with projection operators}
In this section we present Bohmian mechanics in a generalized way,
making extensive use of projection operators, that will allow us
to incorporate operators with discrete spectra. The generalization
agrees with Bohm's original formulation when the beables are
particle position operators but also allows for any choice of
commuting operators ${\hat \xi}_\ell$, $\ell = 1,2,...L$, $[{\hat
\xi}_\ell, {\hat \xi}_{\ell'}]=0$ for the beables. The operators
can have continuous or discrete spectra. Our generalization of
Bohmian Mechanics is not unique. In particular, we have some
freedom to choose how many and which commuting operators we
require to describe the status of all possible measurement
devices.  In this section we will deal only with operators with
continuous spectra and extend the formalism to discrete operators
in the next section.

If ${\hat \xi}_\ell$ has continuous spectra we can express it as
\begin{equation}
{\hat \xi}_\ell = \int d\lambda_\ell \xi_{\ell}(\lambda_\ell)
{\hat P}_\ell(\lambda_\ell)\label{continuous}
\end{equation}
where $\lambda_\ell$ parameterizes the eigenstates of ${\hat
\xi}_\ell$, the integral is taken over the entire range of
$\lambda_\ell$, and $\xi_\ell(\lambda_\ell)$ is the eigenvalue of
${\hat \xi}_\ell$ associated with the eigenstates labeled by
$\lambda_\ell$.
\begin{equation}
{\hat \xi}_\ell |\lambda_\ell,q,\ell>
=\xi_\ell(\lambda_\ell)|\lambda_\ell,q,\ell>
\end{equation}
where $q$ distinguishes states with the same eigenvalue of ${\hat
\xi}_\ell$.  The projection operator for the eigenstates
associated with $\xi_\ell(\lambda_\ell)$ is
\begin{equation}
{\hat
P}_\ell(\lambda_\ell)=\sum_{q}|\lambda_\ell,q,\ell><\lambda_\ell,q,\ell|\label{delta}.
\end{equation}
In this expression the sum over $q$ represents the sum or integral
over states with the same eigenvalue of ${\hat \xi}_\ell$.  The
simplest $\xi_\ell(\lambda_\ell)$ function for an operator with
continuous spectra is $\xi_\ell(\lambda_\ell)=\lambda_\ell$ in
which case $\lambda_\ell$ has units of $\xi_\ell$.  For this case
the projection operator density, ${\hat P}_\ell(\lambda_\ell)$,
takes on the particularly simple form
\begin{equation}
{\hat P}_\ell(\lambda_\ell)= \delta(\lambda - {\hat
\xi}_\ell)\label{delta2}
\end{equation}
For operators with discrete spectra, which we cover in the next
section, we will find it convenient to associate a range of lambda
variables to a single eigenvalue.

Bohmian mechanics, describes the dynamics of a set of
$\lambda_\ell(t)'s$ (that is, $\lambda_1, \lambda_2, \lambda_3
...,$ which we denote collectively by $\Lambda$) and thereby a set
of $\xi_\ell(\lambda_\ell(t))'s$ (which we denote collectively by
$\Xi$) which represents the physical values of these values at
time t, regardless if a measurement is made or not.  The quantum
probability distribution of a particular $\Lambda$ configuration
is conveniently written in terms of the projection operators
\begin{equation}
P({\lambda},|t>)= <t| \prod_{\ell=1}^L {\hat
P}_\ell(\lambda_\ell)|t>\label{prob}
\end{equation}

This is a probability distribution since
\begin{equation}
\int d\lambda_1\int d\lambda_2 ...\int d\lambda_L
f(\xi_1(\lambda_1),
\xi_2(\lambda_2),...\xi_L(\lambda_L))P({\lambda},t) =
<t|f(\hat{\xi}_1, \hat{\xi_2},...\hat{\xi_L})|t>
\end{equation}
  The result is unambiguous since we require
that the $\hat{\xi}$ all commute with each other. The probability
distribution has all the properties required of a classical
probability distribution. The integral of the probability
distribution taken over all of $\Lambda$ space is 1, and the
probability distribution is real and non-negative provided that
all of the projectors in the operator product commute with each
other. The projectors will commute if the ${\hat \xi}_\ell$ all
commute with each other, which is why we made this requirement for
our set of beables.

The quantum state is in general not an eigenstate of the ${\hat
\Xi}$.  It is propagated forwarded in time as in conventional
quantum mechanics
\begin{equation}
i\hbar\frac{d |t>}{d t}= {\hat H}|t>\label{normal}.
\end{equation}
Additionally, the $\Lambda$ configuration is propagated forward in
time with the first order equations
\begin{equation}
\frac{d \lambda_\ell(t)}{dt} = v_\ell(\{\lambda(t)\},t)\label{dyn}
\end{equation}
where the $v_\ell(\{\lambda(t)\},t)$ are chosen so that the
classical probability distribution of the $\lambda$ configuration
of an ensemble of identical experiments
\begin{equation}
P_c({\lambda},t)  = \int d\lambda'_1(0)\int d\lambda'_2(0) ...\int
d\lambda'_L(0)
P_c({\lambda'},0)\prod_{\ell=1}^L\delta(\lambda_\ell-\lambda'_\ell(t,\{\lambda'(0)\})\label{ensemble}
\end{equation}
agrees with the quantum probability distribution at all time
provided they are in agreement at any one time.  If the beables
are chosen such that all measurements are measurements of the
beables, this guarantees that the results of Bohmian mechanics are
consistent with the results of conventional quantum mechanics.

Using Bohmian mechanics Eq.(\ref{dyn}), the time dynamics of the
classical probability distribution is
\begin{equation}
\frac{d  P_c(\{\lambda\},t)}{d t} = -\sum_{\ell=1}^L\frac{\partial
P_c(\{\lambda\},t)v_\ell(\{\lambda\},t)}{\partial
\lambda_\ell}.\label{cdyn}
\end{equation}

Whereas the time derivative of the quantum probability
distribution on the other hand is
\begin{equation}
\frac{\partial  P({\lambda},t)}{\partial t} = \sum_{\ell=1}^L<t|
\left(\prod_{j=1}^{\ell-1}{\hat
P}_j(\lambda_j)\right)\frac{1}{i\hbar}[{\hat
P}_\ell(\lambda_\ell),{\hat H}]\left(\prod_{k=\ell+1}^{L}{\hat
P}_k(\lambda_k)\right)|t>\label{qdyn}.
\end{equation}

Following David Bohm's insight, we note that if the classical and
quantum probability distributions agree at any particular time
then they agree for all time provided that the
$v_\ell(\{\lambda\})$ are chosen so that the two time derivatives
Eqs. (\ref{cdyn}) and (\ref{qdyn}) are equal.  This is what we do
now.

Our goal is to rewrite Eq.(\ref{qdyn}) in the form
\begin{equation}
\frac{\partial  P({\lambda},t)}{\partial t} =
-\sum_{\ell=1}^L\frac{\partial J_\ell({\lambda},t)}{\partial
\lambda_\ell},
\end{equation}
with $J_\ell({\lambda},t)$ real.  If this can be accomplished then
we can set
\begin{equation}
v_\ell(\{\lambda\})=\frac{
J_\ell({\lambda},t)}{P({\lambda},t)}\label{bohm}
\end{equation}
and we will have determined a consistent Bohmian dynamics.
Associating the $\ell$ terms in both expressions we have
\begin{equation}
\frac{\partial J_\ell({\lambda},t)}{\partial \lambda_\ell}=-<t|
\left(\prod_{j=1}^{\ell-1}{\hat
P}_j(\lambda_j)\right)\frac{1}{i\hbar}[{\hat
P}_\ell(\lambda_\ell),{\hat H}]\left(\prod_{k=\ell+1}^{L}{\hat
P}_k(\lambda_k)\right)|t>\label{notreal}.
\end{equation}

Ignoring for the moment the possibility that the right hand side
of Eq.(\ref{notreal}) is not real we write
\begin{equation}
J_\ell({\lambda},t) = <t| \left(\prod_{\ell'=1}^{\ell-1}{\hat
P}_{\ell'}(\lambda_{\ell'}(t))\right)
\hat{J}_\ell(\lambda_\ell(t))\left(\prod_{\ell''=\ell+1}^{L}{\hat
P}_{\ell''}\right)|t>,
\end{equation}
in which
\begin{equation}
\frac{d{\hat J}_\ell(\lambda)}{d \lambda} = -\frac{1}{i\hbar}
[\hat{P}_\ell(\lambda),\hat{H}]
\end{equation}
which is easily integrated to
\begin{equation}
{\hat J}_\ell(\lambda_\ell(t)) =\frac{1}{i\hbar}[
\hat{G}_\ell(\lambda_\ell(t)),{\hat H}] =-\frac{1}{i\hbar}[
\hat{L}_\ell(\lambda_\ell(t)),{\hat H}].
\end{equation}
where
\begin{equation}
\hat{G}_\ell(\lambda_\ell(t)) =\int_{\lambda_\ell(t)}
d\lambda'\hat{P}_\ell(\lambda_\ell')
\end{equation}
and
\begin{equation}
\hat{L}_\ell(\lambda_\ell(t)) =\int^{\lambda_\ell(t)}
d\lambda'\hat{P}_\ell(\lambda_\ell')\label{whatlis}.
\end{equation}
are projection operators for all states greater than or less than
$\lambda_\ell(t)$ respectively.   The current operator, ${\hat
J}_\ell$, can be generalized to periodic beables such as the
position of a bead on a ring with
\begin{equation} {\hat
J}_\ell(\lambda_\ell(t)) =\frac{1}{i\hbar}\int d\lambda'\int
d\lambda''\hat{P}_\ell(\lambda_\ell'){\hat
H}\hat{P}_\ell(\lambda_\ell'')f(\lambda_\ell',\lambda_\ell(t),\lambda_\ell'')
\end{equation}
where $f(\lambda_\ell',\lambda_\ell(t),\lambda_\ell'') = +1 (-1)$
if there is a non-crossing path that goes from $\lambda_\ell'$ to
$\lambda_\ell''$ through $\lambda_\ell(t)$ in the positive
(negative) direction and
$f(\lambda_\ell',\lambda_\ell(t),\lambda_\ell'') = 0$ if there is
no such path.  We will not use this generalization in this paper.

The right hand side of Eq.(\ref{notreal}) is not guaranteed to be
real unless $[[{\hat P}_\ell,{\hat H}],{\hat P}_{\ell'}]=0$ for
$\ell\ne\ell'$ which is true if $[[{\hat \xi}_\ell,{\hat H}],{\hat
\xi}_{\ell'}]=0$ for $\ell\ne\ell'$.  If this is not the case we
can make it real simply by taking the real part
\begin{equation}
J_\ell({\lambda},t) = {\rm Re}\left(<t|
\left(\prod_{\ell'=1}^{\ell-1}\hat{P}_{\ell'}(\lambda_{\ell'}(t))
\right)\hat{J}_\ell(\lambda_\ell(t))\left(\prod_{\ell''=\ell+1}^{L}\hat{P}_{\ell''}\right)|t>\right).
\end{equation}
but this picks out a particular order of the operators for special
treatment.  A more democratic way to guarantee that the current is
real is the symmetric average
\begin{equation}
J_\ell({\lambda},|t>) = <t|S\left \{
\left(\prod_{\ell'=1}^{\ell-1}\hat{P}_{\ell'}(\lambda_{\ell'}(t))
\right)\hat{J}_\ell(\lambda_\ell(t))\left(\prod_{\ell''=\ell+1}^{L}\hat{P}_{\ell''}\right)\right\}|t>
\end{equation}
where $S\{...\}$ implies a symmetric average of all of the
operators inside the braces.  For example if L=3,
\begin{equation}
J_1(\lambda_1,\lambda_2,\lambda_3) = \frac{1}{6}<t|
2\hat{J}_1\hat{P}_2\hat{P}_3 + \hat{P}_2\hat{J}_1\hat{P}_3+
+\hat{P}_3\hat{J}_1\hat{P}_2 +2\hat{P}_2\hat{P}_3\hat{J}_1|t>
\end{equation}
where we have used the fact that the $P$'s commute to combine some
terms. This freedom in the choice of $J_\ell$  is a particular
example of a more general freedom. We can add any function
$Q_\ell$ to the probability current density, $J_\ell$, such that
$\sum_\ell d/d\lambda_\ell Q_\ell=0$. This is the second way that
the dynamics are not unique, (the first being the choice of which
operators to anoint to beable status).

Our final expression for the $\lambda$ dynamics is
\begin{equation}
\frac{d \lambda_\ell(t)}{dt} =
v_\ell(\{\lambda(t)\},t)=\frac{<t|S\left \{
\left(\prod_{\ell'=1}^{\ell-1}\hat{P}_{\ell'}(\lambda_{\ell'}(t))
\right)\hat{J}_\ell(\lambda_\ell(t))\left(\prod_{\ell''=\ell+1}^{L}\hat{P}_{\ell''}\right)\right
\}|t>}{<t| \prod_{\ell=1}^L {\hat
P}_\ell(\lambda_\ell)|t>}.\label{final}
\end{equation}

An equivalent way to write the equations of Bohmian mechanics, is
to use the Heisenberg representation in which the quantum state
does not change with time but any operator $\hat{A}(t)$ depends on
time via
\begin{equation}
\hat{A}(t) =
e^{-\hat{H}t/i\hbar}\hat{A}e^{\hat{H}t/i\hbar}\label{hnormal}.
\end{equation}
The equations of Bohmian Mechanics in the Heisenberg
representation are
\begin{equation}
<0|S\left \{
\left(\prod_{\ell'=1}^{\ell-1}\hat{P}_{\ell'}(\lambda_{\ell'}(t),t)
\right)d\hat{L}_\ell(\lambda_\ell(t),t)\left(\prod_{\ell''=\ell+1}^{L}\hat{P}_{\ell''}\right)\right
\}|0>=0\label{hfinal}
\end{equation}
where
\begin{equation}
d\hat{L}_\ell(\lambda_\ell(t),t) = \frac{\partial
\hat{L}_\ell(\lambda_\ell(t),t)}{\partial t} + \frac{d
\lambda_\ell(t)}{d t}\frac{\partial
\hat{L}_\ell(\lambda_\ell(t),t)}{\partial
\lambda_\ell(t)},\end{equation} in which
\begin{equation} \frac{\partial
\hat{L}_\ell(\lambda_\ell(t),t)}{\partial t}=
\frac{1}{i\hbar}[\hat{L}_\ell(\lambda_\ell(t),t),\hat{H}] =
-\hat{J}_\ell(\lambda_\ell(t),t),
\end{equation}
and
\begin{equation}
\frac{\partial \hat{L}_\ell(\lambda_\ell(t),t)}{\partial
\lambda_\ell(t)}= \hat{P}_\ell(\lambda_\ell(t),t).
\end{equation}

In the Schrodinger representation, since the quantum state changes
with time, one is tempted to think of it as a dynamic variable
just like the beables and wonder why the beables depend on the
quantum state but not the other way around.  In the Heisenberg
representation the quantum state is not a dynamic variable so this
apparent asymmetry does not arise.

If an ensemble of identical experiments are performed, in which
the $\Lambda$ configurations at a particular time for each
experiment are taken at random from the quantum distribution
Eq.(\ref{prob}) then if the $\Lambda$ configurations are
propagated forwards and backwards in time via Eq.(\ref{final}) or
Eq.(\ref{hfinal}) then for all other times the probability
distribution of the $\lambda$ configurations over the ensemble
Eq.(\ref{ensemble}) will equal the quantum distribution. Provided
that the set of anointed operators is sufficient to describe the
status of all measurement devices, Bohmian mechanics will agree
with all results of conventional quantum mechanics without
resorting to wavefunction collapse or some other alternative to
the Schrodinger dynamics to describe the measurement process as is
done in the orthodox Copenhagen interpretation.  Bohmian mechanics
replaces this with the mystery of how to explain why the classical
and quantum probability distributions should agree at any time at
all\cite{goldstein}.

If all of the $\lambda_\ell$'s correspond to position coordinates
in nonrelativistic quantum mechanics then the conventional form of
Bohmian mechanics is recovered from Eq.(\ref{final}) or
Eq.(\ref{hfinal}). For then the current operators are
\begin{equation}
{\hat J}_\ell(\lambda_\ell) = \frac{1}{2}[\frac{{\hat
p}_\ell}{m_\ell},\delta(\lambda_\ell - {\hat x}_\ell)]_+.
\end{equation}
This is the current operator in conventional Bohmian mechanics so
the equivalence with the traditional formalism is proved.  The
present formalism is more flexible than the traditional formalism
and can be easily generalized to account for beable operators with
discrete spectra, which is what we consider in the next section.

\section{fitting the discrete square peg into the continuous round hole}
In this section we incorporate beables derived from operators with
discrete spectra into Bohmian mechanics. Doing so, we are
immediately faced with the question of how to retain determinism
which Bell sacrificed with regret.   The problem is most apparent
when the initial quantum state is an eigenstate of each $\xi_\ell$
for then the initial $\Xi$ configuration is uniquely determined so
there appears to be no room for any randomness initially.  The
quantum time dynamics will immediately make the quantum state a
superposition of $\Xi$ states so the classical probability
distribution must develop some spreading to agree with its quantum
counterpart. Since there is no randomness in the initial
conditions, it appears that randomness
 must enter through the dynamics and therefore determinism
must be sacrificed.  Note that this is only a problem if all of
the $\xi_\ell$ operators have discrete spectra. If even just one
of the $\xi_\ell$ operators have continuous spectra then that is
enough to make the initial state not unique so there is the
possibility of a deterministic dynamics producing the correct
future probability distributions. Strictly speaking the same
problem can arise for operators with continuous spectra but for
continuous spectra one can wiggle one's way out by asserting that
in any actual experiment the quantum state is never exactly in an
eigenstate of all of the operators. There is always some spreading
for whatever reason.

A way to retain deterministic dynamics with discrete operators is
to assign finite ranges of $\lambda_\ell$ to the same eigenvalue
of ${\hat \xi}_\ell$. $\xi_\ell$ sits still when $\lambda_\ell$ is
moving smoothly through a region that corresponds to the same
eigenvalue and $\xi_\ell$ makes a sudden hop to a new eigenvalue
when $\lambda_\ell$ smoothly moves from one eigenvalue region to
another. The problem of determinism is solved by this devise since
there are now many $\Lambda$ configurations corresponding to the
same physical state. It might appear that for operators with
discrete spectra, $\lambda_\ell$ can legitimately be called a
hidden variable since the particular value of $\lambda_\ell$
inside an eigenvalue range is unobservable and therefore "hidden".
But $\lambda_\ell$ is observable at those times when the
$\xi_\ell$ hops to a new value since there is a unique value of
$\lambda_\ell$ for each transition.  Therefore $\lambda_\ell$
encodes observable information about previous transitions and the
times that they occurred. We now construct the explicit Bohmian
mechanics for the discrete case.

If ${\hat \xi}_\ell$ has discrete spectra we can express it as
\begin{equation}
{\hat \xi}_\ell = \sum_n \xi_{\ell n} {\hat
P}_\ell(n)\label{discrete}
\end{equation}
where $\xi_{\ell n}$ is the nth eigenvalue of ${\hat \xi}_\ell$
and the projection operator  is
\begin{equation}
{\hat P}_\ell(n) = \sum_{q}|n,q,\ell><n,q,\ell|
\end{equation}
,where as in the continuous case, the sum over $q$ represents the
sum or integral over states with the same eigenvalue of ${\hat
\xi}_\ell$.

We now seek to express Eq.(\ref{discrete}) in the continuous form
Eq.(\ref{continuous}) and define an appropriate
$\xi_\ell(\lambda_\ell)$ function and projection operator
$\hat{P}_\ell(\lambda_\ell)$ so that we may carry over
Eq.(\ref{final}) or Eq.(\ref{hfinal}) unchanged for the dynamics.
There are many ways to do this.  Technically,
$\theta_\ell(\lambda_\ell) = \lambda_\ell$ and ${\hat
P}_\ell(\lambda) = \delta(\lambda - {\hat \xi}_\ell)$ as in the
continuous case does the job.  But this leads to zero probability
for $\lambda_\ell$ not equal to an eigenvalue. We can correct for
this by smearing out the delta function over a range of
$\lambda_\ell$ so that ${\hat P}_\ell(\lambda_\ell)$ is not zero
between eigenvalues and a range of $\lambda_\ell$ corresponds to
the same state.  There are innumerable ways to parameterize
$\xi_\ell$ to achieve this. Here is one way. Define the $\xi_\ell$
function
\begin{equation}
\xi_\ell(\lambda_\ell) = \xi_{\ell,n}\;\;\;
n=n(\lambda_\ell)\label{fix}
\end{equation}
where $\lambda_\ell$ has no units and $n(\lambda_\ell)$ is the
closest integer to $\lambda_\ell$. With this
$\xi_\ell(\lambda_\ell)$ function we achieve agreement between
Eq.(\ref{discrete}) and Eq.{\ref{continuous}) using
\begin{equation}
{\hat P}_\ell(\lambda_\ell) ={\hat
P}_\ell(n(\lambda_\ell)).\label{integer}
\end{equation}
where ${\hat P}_\ell(n(\lambda_\ell))=0$ if there is no eigenvalue
associated with the integer $n(\lambda_\ell)$.  For discrete
spectra with integer eigenvalues we can use the explicit forms
\begin{eqnarray}
\hat{G}_\ell\left(\lambda_\ell(t)\right) &=&\left(n_\ell(t) + 1/2
-\lambda_\ell(t)\right)\hat{P}_\ell\left(n_\ell(t)\right)+
\sum_{j=n_\ell(t)+1}^\infty\hat{P}_\ell(j)\nonumber\\
\hat{L}_\ell\left(\lambda_\ell(t)\right)
&=&\left(\lambda_\ell(t)-n_\ell(t) +
1/2\right)\hat{P}_\ell\left(n_\ell(t)\right)+
\sum_{j=-\infty}^{n_\ell(t)-1}\hat{P}_\ell(j),
\end{eqnarray}
in the expressions for the current operator ${\hat J}_\ell$.

Using these definitions in the velocity expression
Eq.(\ref{final}) or Eq.(\ref{hfinal}) and keeping in mind that the
physical values of $\xi_\ell$ with discrete spectra are determined
by Eq.(\ref{fix}) we have defined a deterministic Bohmian
mechanics for operators with discrete and continuous spectra. The
initial $\Lambda$ configuration is taken from the quantum
probability distribution Eq.(\ref{prob}) with Eq.(\ref{integer})
used for projectors for discrete operators. This means that the
particular $\lambda_\ell$ for a given $\xi_{\ell, n}$ is chosen at
random from a distribution spread uniformly from $n-1/2$ to $n+
1/2$. We are free to order the eigenvalues along the
$\lambda_\ell$ line any way we like, a freedom that also exists in
the continuous case.  This is the third way that the dynamics are
not unique. The choice of ordering of the eigenvalues in $\Lambda$
space for each beable operator profoundly effects the dynamics
since, for example a system in eigenvalue state 2 can only get to
eigenvalue state 4 by first passing through eigenvalue state 3.
For some systems the Hamiltonian determines a particular order,
but perhaps there are Hamiltonians with transition elements
between eigenvalues that are far apart in $\Lambda$ space, for any
ordering that one chooses.  This is a possible disadvantage of
this method which is not shared by the stochastic Bell scheme.

In the next section we show how the formalism works for the simple
example of Bohmian mechanics with only one beable and also present
a visualization of Bohmian mechanics that allows us to dispense
with the auxiliary $\lambda_\ell$ variables altogether.

\section{The example of one beable and a visualization of Bohmian mechanics}
Bohmian mechanics is integrable for the case in which there is
only one operator promoted to beable status.  For in that case
\begin{equation}
<0|d\hat{L}(\lambda(t),t)|0> =d<0|\hat{L}(\lambda(t),t)|0>=0.
\end{equation}
so the solution is
\begin{equation}
<0|\hat{L}(\lambda(t),t)|0> = <t|\hat{L}(\lambda(t))|t>=L_0.
\end{equation}
The integration constant, $L_0$ is uniformly distributed between
zero and one. This equation has the following visual
interpretation. Consider a line that goes from zero to one. The
integration constant, $L_0$ sits immovable on the line. Associate
the portion of the line from zero to $<t|\hat{P}_a|t>$ with beable
value $a$, the portion of the line from $<t|\hat{P}_a|t>$ to
$<t|\hat{P}_a|t>+ <t|\hat{P}_b|t>$ with beable value $b$, the
portion of the line from $<t|\hat{P}_a|t>+ <t|\hat{P}_b|t>$ to
$<t|\hat{P}_a|t>+ <t|\hat{P}_b|t>+ <t|\hat{P}_c|t>$ with beable
value $c$. These boundaries change with time. The value of the
beable at any time is the value associated with the portion of the
line that $L_0$ sits on at that time. As an example consider a two
state case in which an operator $\hat{\xi}$, with eigenvalues $\pm
1$ and projectors $\hat{P}_\pm$ is the beable (we have chosen a
slightly different parameterization of the eigenvalues than we did
in the previous section to take advantage of the symmetry in the
two state case) . The Bohmian dynamics dictate that $\xi=-1$ for
$<t|\hat{P}_-|t>$ $>$ $L_0$ and $\xi=+1$ for $L_0$ $>$
$<t|\hat{P}_-|t>$.  These results can be combined into the
equation of motion
\begin{equation}
\xi(t) = \rm{sign}(L_0 - <t|\hat{P}_-|t>)
\end{equation}
or using $<t|\hat{\xi}|t> = 1-2<t|\hat{P}_-|t>$
\begin{equation}
\xi(t) = \rm{sign}(<t| \hat{\xi}|t> - \xi_0)\label{twostate1}.
\end{equation}
where $\xi_0 = 1-2L_0$ which is uniformly distributed from -1 to
1. Note that $\xi_0$ encodes observable information about the
times $t_j$ that $\xi$ changes its state via $<t_j|\hat{\xi}|t_j>
= \xi_0$ so it is not "hidden", although it is uncontrollable.
Also, the average value of $\xi(t)$ over all $\xi_0$ agrees with
the quantum expectation value
\begin{equation}
\int_{-1}^{+1}\xi(t)P(\xi_0)d \xi_0 =<t| \hat{\xi}|t>
\end{equation}
as is required for Bohmian mechanics to be consistent with quantum
mechanics.

The exact solution for one beable suggests a visual interpretation
of Bohmian Mechanics for any number of beables that allows us to
dispense with $\lambda_\ell$ as we were able to do for the one
beable case. For n beables consider an n dimensional space of area
1 with fixed boundaries. The space is divided into several
n-dimensional bubbles.  Each bubble corresponds to a particular
$\Xi$ configuration and the volume of each bubble is the quantum
probability of that configuration. Since the probabilities change
with time, the bubbles are continuously contracting and expanding
against each other.  An immovable point is chosen at random in the
n-dimensional space. The physical $\Xi$ configuration at time t is
the $\Xi$ configuration corresponding to the bubble enclosing the
immovable point at time t.

\section{summary}
In this paper we have generalized Bohmian mechanics so that it can
incorporate beables associated with an arbitrary set of commuting
continuous and discrete operators. The equations are deterministic
and time reversible and agree with Bohm's original formulation for
the case of continuous position operators.  The simple case of
only one beable is presented and the solution suggests an
intriguing visualization of Bohmian mechanics.

We acknowledge support from Research Corporation grant CC5326.
\end{document}